\documentclass{lmcs} %%% last changed 2014-08-20


\usepackage{graphicx}
\usepackage{amssymb}
\usepackage{url}
\usepackage{listings}
\theoremstyle{plain} %\crefname{satz}{Satz}{S\"atze}
\begin{document}

\title{Formalization of the Axiom of Choice and its Equivalent Theorems}
%\titlecomment{{\lsuper*}OPTIONAL comment concerning the title, \eg,
%  if a variant or an extended abstract of the paper has appeared elsewhere.}

\author[Tianyu Sun]{Tianyu Sun}	%required
\address{Beijing Key Laboratory of Space-ground Interconnection and Convergence, Beijing University of Posts and Telecommunications, Beijing, China}	%required
\email{\{stycyj,wsyu\}@bupt.edu.cn}  %optional
%\thanks{thanks 1, optional.}	%optional

\author[Wensheng Yu]{Wensheng Yu}	%optional
%\address{Beijing Key Laboratory of Space-ground Interconnection and Convergence, Beijing University of Posts and Telecommunications, Beijing 100876, China.}	%optional
%\email{wsyu@bupt.edu.cn}  %optional
%\thanks{This work was supported by National Natural Science Foundation (NNSF) of China under Grant 61571064.}	%optional

%\author[C.~Name3]{Carla Name3}	%optional
%\address{address 3}	%optional
%\urladdr{name3@url3\quad\rm{(optionally, a web-page can be specified)}}  %optional
%\thanks{thanks 3, optional.}	%optional

%% etc.

%% required for running head on odd and even pages, use suitable
%% abbreviations in case of long titles and many authors:

%%%%%%%%%%%%%%%%%%%%%%%%%%%%%%%%%%%%%%%%%%%%%%%%%%%%%%%%%%%%%%%%%%%%%%%%%%%

%% the abstract has to PRECEDE the command \maketitle:
%% be sure not to issue the \maketitle command twice!

\begin{abstract}
  \noindent In this paper, we describe the formalization of the axiom of choice and several of its famous equivalent theorems in Morse-Kelley set theory. These theorems include Tukey's lemma, the Hausdorff maximal principle, the maximal principle, Zermelo's postulate, Zorn's lemma and the well-ordering theorem. We prove the above theorems by the axiom of choice in turn, and finally prove the axiom of choice by Zermelo's postulate and the well-ordering theorem, thus completing the cyclic proof of the equivalence between them. The proofs are checked formally using the Coq proof assistant in which Morse-Kelley set theory is formalized. The whole process of formal proof demonstrates that the Coq-based machine proving of mathematics theorem is highly reliable and rigorous. The formal work of this paper is enough for most applications, especially in set theory, topology and algebra.
\end{abstract}

\maketitle

%% start the paper here:
\section{Introduction}
\label{sec:1}
In the late 19th century and early 20th century, some paradoxes of naive set theory are discovered, which made the research of axiomatic set theory necessary \cite{BF58,F66,FBL73,J03,K55}. The axiomatic set theory need to offer a series of axioms that sets should meet, and the properties of sets are studied based on these axioms. The set theory applies ZFC as its axiom system generally, which is composed of Zermelo-Fraenkel set theory and the axiom of choice (AC). Other common axiomatic set theory includes von Neumann-Bernays-G\"{o}del set theory (NBG) and Morse-Kelley set theory (MK). \cite{BF58,F66,FBL73,H67,J73,J03,K55,M65,W49,Z10,Z35}.

MK was first proposed by Wang Hao \cite{W49} in 1949, and was formally published in Kelley's $General$ $Topology$ \cite{K55} in 1955. Morse later presented his own version in 1965 \cite{M65}. MK is a variant of ZF system which improved by Skolem and Morse, and it is more close to NBG. Meanwhile, it is designed to give quickly and naturally a foundation for mathematics which is free from the more obvious paradoxes. The ordinal, cardinal numbers and non-negative integers are constructed and Peano's postulates are proved as theorems. Furthermore, the real numbers can be constructed from integers by use of the axiom of infinity and two facts: ``the class of integers is a set'' and ``it is possible to define a function on the integers by induction'' \cite{K55,L51}. In this axiom system, we admit classes as basic objects, like NBG. In addition, a finite axiom system is abandoned and the development is based on eight axioms and one axiom scheme (that is, all statements of a certain prescribed form are accepted as axioms) \cite{K55}. Thus MK cannot be finitely axiomatized and it is strictly stronger than both NBG and ZF. In fact, NBG and ZFC can be proved consistent in MK. Monk, Rubin and Mendelson submit that MK does what is expected of a set theory while being less cumbersome than ZFC and NBG \cite{BF58,F66,FBL73,H67,J03,K55}.

AC is an axiom about the existence of mapping in set theory. It was first proposed by Zermelo in 1904 and used for proving the well-ordering theorem \cite{Z10}. In 1935, Zorn proved a famous principle in algebra with AC, which was called Zorn's lemma \cite{Z35}. AC has a very important role in modern mathematics and has very close ties with many profound mathematics conclusions. Without AC, we don't even know if two sets can compare their element numbers with each other, if the product of a family of nonempty set is empty, if the liner space must have a group of bases, if any family of compact space must be compact and so on \cite{BF58,J73}. AC has a lot of equivalent forms. The most famous ones including Tukey's lemma, the Hausdorff maximal principle, the maximal principle, Zermelo's postulate, Zorn's lemma, the well-ordering theorem and so on.

In recent years, with the rapid development of computer science, especially the emergence of interactive theorem proving tools Coq \cite{BC04,T18}, Isabelle/HOL \cite{NPW02} and so on, the formalization of mathematics has made great progress \cite{A18,G08,GAABC13,H08,W08}. In 2005, the international computer experts Gonthier and Werner offered the machine proving of the famous ``four-color theorem'' using Coq successfully\cite{G08}. After six years of hard work, Gonthier completed the machine verification of the ``odd order theorem'' in 2012 \cite{GAABC13}, which made Coq more and more popular in academia. Wiedijk pointed out that relevant research group around the world have completed or plan to complete the formal proof of 100 famous mathematical theorems including G\"{o}del incompleteness theorem, Jordan curve theorem, Prime number theorem, Fermat last theorem and so on \cite{W08}.

On the basis of the ``axiomatic set theory'' formal system which is developed according to MK, we formalize AC and several of well-known theorems. Furthermore, we implement the formal proof of the equivalence between them in this paper. These theorems include Tukey's lemma, the Hausdorff maximal principle, the maximal principle, Zermelo's postulate, Zorn's lemma and the well-ordering theorem. It should be noted that AC and its equivalent theorems have been required for many fields of modern mathematics. Their traditional proving process can be found in numerous references respectively, such as \cite{BF58,F66,FBL73,J73,J03,K55,Z10,Z35}.

The innovation and contribution of this paper is as follows. First, we formally implement the cyclic proof of AC and the six famous theorems. This proves the equivalence between AC and them. The cyclic proof is first formalized to our knowledge. The relationship between these theorems are systematically discussed in this paper. Second, the formal proof of this paper has many advantages over traditional proofs. The process of formal proof is more complete and every detail is verified. This makes up for the many proof details in traditional proofs. At the same time, this is also the verification of traditional proof. In addition, we have some innovations of formal methods in the proof process. The proofs are checked formally using the Coq proof assistant, which fully embodies the Coq-based mathematics theorem machine proving system are readable, interactive and intelligent. The machine proving progress is rigorous and reliable. Third, the formalization of this paper is based on the ``axiomatic set theory'' formal system which we developed. It avoids the problem of type confusion and some paradoxes compared with the naive set theory \footnote{The naive set theory in Coq standard library are available at \url{https://coq.inria.fr/distrib/current/stdlib/Coq.Sets.Ensembles.html/}} in Coq standard library. It establishes a foundation for further development. Forth, there are many important applications of the formalization in this paper. We can discuss the formalization of more equivalent theorems of AC, and we can prove many famous theorems including Tychonoff's theorem in topology, the Banach-Tarski paradox in measure theory, the Nielsen-Schreier theorem in algebra, the Hahn-Banach theorem in functional analysis and so on. Based on the ``axiomatic set theory'' formal system, we can construct various mathematical theories, such as topology, measure theory, abstract algebra and so on.

The organization of the paper is as follows: in section\,2, we introduce the ``axiomatic set theory'' formal system which we developed and some modifications in this paper. In section\,3, we present some definitions needed in the proof. In this paper, we give the exact Coq description code after the definitions and theorems. In section\,4, we discuss the cyclic proof of AC and several of its famous equivalent theorems in detail. It should be pointed out that all the proof in this paper are verified in Coq. Section\,5 is dedicated to related work. Finally, in section\,6, we draw conclusions and discuss some potential further work.

\section{Axiomatic Set Theory}
\label{sec:2}
The ``axiomatic set theory'' formal system is built on the basis of MK. It is designed to give quickly and naturally a foundation for mathematics. Compared with naive set theory in Coq standard library, the system has many advantages. First of all, it avoids the more obvious paradoxes of naive set theory. Secondly, there is no type difference between sets and members in this system. The universe of discourse consists of classes. Classes which are members of other classes are called sets. A class which is not a set is a proper class. These help the system avoid the problem of nested type mismatch in Coq formalization. Thirdly, the system is more complete for the Coq formalization of set theory. The system has a wide range of applications in many aspects, and the work of this paper is one of its important applications. In addition, topology and abstract algebra can be quickly formalized on the basis of the system.

We have completed the Coq formalization of the system, including eight axioms, one axiom schema, 62 definitions and 148 corollaries or theorems. The complete source is available online:

\vskip 0.2cm
{\noindent \url{https://github.com/styzystyzy/Axiomatic_Set_Theory/}}
\vskip 0.2cm

Next we introduce some basic definitions and the axiom system of the formal system. For the sake of space, we only list some important contents which are used in this paper to demonstrate the formalization of the system.

In the first, we define the `Class' which can describe the type of sets and members in the system. We choose to let it live in Type, which is the topmost Sort in Coq. Formally, the definition is:

\begin{lstlisting}
Parameter Class : Type.
\end{lstlisting}

In this system, we admit the equality `=' and some basic logical constants, including the negation `$\sim$', the conjunction `$/\backslash$', the disjunction `$\backslash/$', universal quantification `$\forall$', existential quantification `$\exists$' and so on. We also recognize some basic logical properties\footnote{We use axioms and lemmas from the library $\bf{Coq.Logic.Classical}$ in the system.}. In addition, there are two primitive constants besides `=' and the other logical constants. The first of those is `$\in$', which is read `is a member of' or `belongs to'. Since we do not distinguish the type of sets and members, the formal statement of `$\in$' is as follows.

\begin{lstlisting}
Parameter In : Class -> Class -> Prop.
\end{lstlisting}

Notations can be introduced to ease reading and writing of specifications. This also allows us to stay close to the way mathematicians would write. Moreover, we can also define precedence levels and associativity rules of notations in Coq. The symbol of `$\in$' is defined in Coq as follows:

\begin{lstlisting}
Notation "x $\in$ y" := (In x y) (at level 10).
\end{lstlisting}

The second constant is the classifier `$\{ \cdots : \cdots \}$' and is read `the class of all $\cdots$ such that $\cdots$'. For example, $\{ x : x\in y \}$ is a classifier. The first blank in the classifier constant is to be occupied by a variable which represents the member of the classifier. The second blank is to be occupied by a formula. It should be noted that the formula here can be any property, including incorrect property. The classifier is just a class, and we don't know if there are members in it.

The formalization of the classifier is divided into two cases. If the member is a single item, then $Classifier$ is used. If the member is a ordered pair, then $Classifier\_\,P$ is used. The definition of $Classifier\_\,P$ is to avoid excessive notation. We agree that $\{ (x,y) : \cdots \}$ is to be identical with $\{ u : $ for some $x$, some $y$, $u = (x,y)$ and $\cdots \}$. The formal statement and symbolic representation of the classifier are as follows:

\begin{lstlisting}
Parameter Classifier : (Class -> Prop) -> Class.
Parameter Classifier_P : (Class -> Class -> Prop) -> Class.
Notation "\{ P \}" := (Classifier P) (at level 0).
Notation "\{\ P \}\" := (Classifier_P P) (at level 0).
\end{lstlisting}

Next we introduce some basic definitions in the system. We list their Coq code, mathematical definition and notation in formalization, as shown in Table\,1.

\begin{table}[ht]
  \label{tab:1}
  \renewcommand\arraystretch{1.3}
  \centering
  \caption{Some important definitions of the system}
  \begin{tabular}{| c | c | c |}
    \hline
   Coq Code             & Mathematical Definition                & Notation \\ \hline
   Ensemble x           & The class $x$ is a set                 & none  \\ \hline
   Union x y            & The union of x and y                   & $x\cup y$    \\ \hline
   Intersection x y     & The intersection of x and y            & $x\cap y$    \\ \hline
   Setminus x y         & The difference of $x$ and $y$          & $x\sim y$    \\ \hline
   Inequality x y       & $x = y$ is not true                    & $x\neq y$    \\ \hline
   Empty                & The Empty is the void class, or zero   & $\emptyset$  \\ \hline
   Full                 & The Full is the universe               & $\mathcal{U}$   \\ \hline
   Element\_U x         & The union of members of $x$            & $\bigcup x$     \\ \hline
   Element\_I x         & The intersection of members of $x$     & $\bigcap x$     \\ \hline
   Subclass x y      & $x$ is contained in $y$, or $y$ contains $x$  & $x\subset y$ \\ \hline
   ProperSubclass x y   & $y$ properly contains $x$          & $x\subsetneq y$     \\ \hline
   PowerClass x         & The power set of $x$, denoted $2^{x}$  & $pow( x )$      \\ \hline
   Singleton x          & The singleton $\{x\}$ of a class $x$   & $[ x ]$      \\ \hline
   Unordered x y   & The class $\{ x\,y \}$ is an unordered pair & $[ x | y ]$   \\ \hline
   Ordered x y     &  The class $(x,y)$ is an ordered pair   & $[ x , y ]$      \\ \hline
   First z         & The first coordinate of the ordered pair $z$& $fst( z )$   \\ \hline
   Second z       & The second coordinate of the ordered pair $z$& $snd( z )$   \\ \hline
   Function f           & The class $f$ is a function            & none \\ \hline
   Domain f             & The domain of function $f$             & $dom( f )$  \\ \hline
   Range f              & The range of function $f$              & $ran(f)$   \\ \hline
   Value f x            & The value of $f$ at $x$, denoted $f(x)$ & $f [ x ]$    \\ \hline
   Cartesian x y        & The Cartesian product of $x$ and $y$   & $x\times y$  \\ \hline
   Rrelation x r y   & $x$ is $r$-related to $y$, denoted $x$ $r$ $y$  & none  \\ \hline
   Finite x             & The class $x$ is finite                & none  \\ \hline
  \end{tabular}
\end{table}

Finally, we describe the formalization of the axiom system which consists of one axiom scheme and eight axioms. The classification axiom scheme is very important in this system and some obvious paradoxes in set theory can be avoided by it. A precise statement of the classification axiom scheme requires a description of formulate. It is agreed that:

\begin{enumerate}
  \item[(a)] The result of replacing `$\alpha$' and `$\beta$' by variables is, for each of the following, a formula.
      \vskip0.2cm
      \begin{center}
       $  \alpha = \beta  \quad\quad  \alpha\in \beta  $
      \end{center}
      \vskip0.2cm
  \item[(b)] The result of replacing `$\alpha$' and `$\beta$' by variables and `$A$' and `$B$' by formulae is, for each of the following, a formula.
      \vskip0.2cm
      \begin{center}
      if $A$, then $B$ \quad $A$ iff $B$ \quad it is false that $A$\\
      $A$ and $B$ \quad $A$ or $B$ \\
      for every $\alpha$, $A$ \quad for some $\alpha$, $A$\\
      $\beta\in\{\alpha: A\}$ \quad $\{\alpha: A\}\in\beta$ \quad $\{\alpha: A\}\in\{\beta: B\}$
      \end{center}
      \vskip0.2cm
      Formulaes are constructed recursively, beginning with the primitive formulae of (a) and proceeding via the constructions permitted by (b).
\end{enumerate}

In the following '$\alpha$' and '$\beta$' are replaced by variables. The formulae $F$ of '$F(\alpha)$' is constructed by the above method. And '$F(\beta)$' is represented by the formulae obtained from '$F(\alpha)$' by replacing each occurrence of the variable which replaced $\alpha$ by the variable which replaced $\beta$. The formulae $F$ may contain parameters which are either sets or proper classes. More consequentially, the quantified variables in $F$ may range over all classes and not just over all sets. But the members of $\{\alpha : F(\alpha)\}$ are exactly those sets such that $F$ comes out true. The specific description of the axiom is as follows:

\vskip 0.25cm
{\noindent
{\bf Classification axiom-scheme}\quad{For each $\beta$, $\beta \in \{\,\alpha : F(\alpha)\,\}$ if and only if $\beta$ is a set and $F(\beta)$. }}
\vskip 0.25cm

For the two cases of the classifier, the formalization of the axiom-schema is also divided into two cases. Nevertheless, the principles of two cases are consistent. In the second case, we often encounter situations where the input value is a single variable in the Coq proof. However, the input variable of $Axiom\_SchemP$ is required to be an ordered pair. Therefore, we need to add a property to make the single variable into an ordered pair when there is a single variable belongs to the classifier of ordered pairs.

\begin{lstlisting}
Axiom Axiom_Scheme : $\forall$ (b: Class) (F: Class -> Prop),
  b $\in$ \{ F \} <-> Ensemble b /\ (F b).
Axiom Axiom_SchemeP : $\forall$ (a b: Class)(F: Class -> Class -> Prop),
  [a,b] $\in$ \{\ F \}\ <-> Ensemble [a,b] /\ (F a b).
Axiom Property_P : $\forall$ (z: Class) (F: Class -> Class -> Prop),
  z $\in$ \{\ F \}\ -> ($\exists$ a b, z = [a,b]) /\ z $\in$ \{\ F \}\.
\end{lstlisting}

The above formalization is defined on all properties $F$: Class $\rightarrow$ Prop. It is a second-order logic and is stronger than Kelley's first-order logic. In this system, we only apply the axiom to formulaes constructed by the above method.

In addition to the classification axiom scheme, there are eight axioms in the system. The development of the whole axiom system is based on these eight axioms and the classification axiom scheme. The Coq statement of these axioms are as follows:

\begin{lstlisting}
Axiom Axiom_Extent : $\forall$ x y, x = y <-> ($\forall$ z, z $\in$ x <-> z $\in$ y).
Axiom Axiom_Subsets : $\forall$ x: Class,
  Ensemble x -> $\exists$ y, Ensemble y /\ ($\forall$ z, z $\subset$ x -> z $\in$ y).
Axiom Axiom_Union : $\forall$ x y, Ensemble x /\ Ensemble y -> Ensemble (x $\cup$ y).
Axiom Axiom_Substitution : $\forall$ f: Class,
  Function f -> Ensemble dom(f) -> Ensemble ran(f).
Axiom Axiom_Amalgamation : $\forall$ x, Ensemble x -> Ensemble ($\bigcup$ x).
Axiom Axiom_Regularity : $\forall$ x, x $\neq$ $\emptyset$ -> $\exists$ y, y $\in$ x /\ (x $\cap$ y) = Empty.
Axiom Axiom_Infinity : $\exists$ y: Class,
  Ensemble y /\ $\emptyset$ $\in$ y /\ ($\forall$ x, x $\in$ y -> (x $\cup$ [x]) $\in$ y).
Axiom Axiom_Choice : $\exists$ c, ChoiceFunction c /\ dom(c) = $\mathcal{U}$ $\sim$ [$\emptyset$].
\end{lstlisting}

For the completeness and independence of the formalization, the set theory used in this paper is a simplified and modified version of the ``axiomatic set theory'' formal system. It contains the complete axiom system, as well as approximately 50 definitions and 70 theorems of the above system.

The modifications are as follows. Firstly, for the independence of the formalization of AC and its related theorems, we redefine the choice function in section\,3 and restate AC in section\,4.1. The choice function in section\,3 is more specific and there is no contradiction between the two descriptions of AC. Secondly, the well order in the well-ordering theorem of this paper is non-strict and it is strict in the system. The distinction between strict and non-strict well orders is often ignored since they are easily interconvertible. In addition, we also add some basic definitions in section\,3, such as partial orders, total orders and so on. Thirdly, partial orders, total orders, and well orders of this paper are all limited to a set, but the definition of well orders in the system is more general and does not have this limitation. We can derive the former from latter. Therefore, there is no contradiction between them.

\section{Basic Definitions and Properties}
\label{sec:3}
Before explaining the proof of the equivalence between AC and theorems, we need to introduce some basic definitions and properties involved in the proof. The specific content and the Coq implementation of them are as follows.

First of all, we give the definition of the choice function. The formalization of AC will be defined by the choice function.

\begin{defi}[Choice Function]
$\varepsilon$ is a choice function of set $X$ if and only if $\varepsilon$ is a function, the domain of $\varepsilon$ is $2^X \sim\{\phi\}$, the range of $\varepsilon$ is a subclass of $X$ and $\varepsilon(A)\in A$ for each member $A$ of domain $\varepsilon$.
\end{defi}

\begin{lstlisting}
Definition Choice_Function ($\varepsilon$ X: Class) : Prop :=
  Function $\varepsilon$ /\ dom($\varepsilon$) = pow(X) $\sim$ [$\emptyset$] /\ ran($\varepsilon$) $\subset$ X /\
  ($\forall$ A, A $\in$ dom($\varepsilon$) -> $\varepsilon$[A] $\in$ A).
\end{lstlisting}

The following definitions are very important, and they are used in Tukey's lemma, the Hausdorff maximal principle and so on.

\begin{defi}[Maximal (Minimal) Member]
$F$ is a maximal (minimal) member of $f$ iff no member of $f$ properly contains $F$ (no member of $f$ is properly contained in $F$).
\end{defi}

When $f$ is equal to empty, $f$ has no maximal (minimal) member. Thus we add the condition $f \neq \emptyset$ when we formalize the maximal (minimal) member. The condition is very important in proving the existence of maximal elements in a set. At the same time, it eliminates many unnecessary discussions.

\begin{lstlisting}
Definition MaxMember (F f: Class) : Prop :=
  f $\neq$ $\emptyset$ -> F $\in$ f /\ ($\forall$ E, E $\in$ f -> $\sim$ (F $\subsetneq$ E)).
Definition MinMember (F f: Class) : Prop :=
  f $\neq$ $\emptyset$ -> F $\in$ f /\ ($\forall$ E, E $\in$ f -> $\sim$ (E $\subsetneq$ F)).
\end{lstlisting}

The following is the definition of nest. It will be used in the description of the Hausdorff maximal principle. The specific description and Coq formalization of it are as follows:

\begin{defi}[Nest]
$n$ is a nest if and only if, whenever $x$ and $y$ are members of $n$, then either $x\subset y$ or $y\subset x$.
\end{defi}

\begin{lstlisting}
Definition Nest n : Prop := $\forall$ x y, x $\in$ n /\ y $\in$ n -> x $\subset$ y \/ y $\subset$ x.
\end{lstlisting}

Next we present the definition of the set of finite character. We will use this definition in the description of Tukey's lemma.

\begin{defi}[Set of Finite Character]
A set $f$ is of finite character provided it has the following properties:

\vskip 0.1cm
\par\setlength\parindent{0.5em} (1) For each $F\in f$, every finite subclass of $F$ belongs to $f$.

\vskip 0.1cm
\par\setlength\parindent{0.5em} (2) If every finite subclass of a given set $F$ belongs to $f$, then $F\in f$.
\end{defi}

The formalization of the set of finite character involves the definition of the finite. We use the definition of finite in the ``axiomatic set theory'' formal system. It is based on the definition of integers and cardinal numbers. Its Coq formal statement is as follows.

\begin{lstlisting}
Definition Finite_Char (f: Class) : Prop :=
  Ensemble f /\ ($\forall$ F, F $\in$ f -> ($\forall$ z, z $\subset$ F /\ Finite z -> z $\in$ f)) /\
  ($\forall$ F, Ensemble F /\ ($\forall$ z, z $\subset$ F /\ Finite z -> z $\in$ f) -> F $\in$ f).
\end{lstlisting}

According to definitions of the set of finite character and the nest, we can prove some properties. They will be used repeatedly in subsequent proofs. The details are as follows:

If $f$ is a non-empty set of finite character, then it has the following properties:
(1) For each $A\in f$, every finite subclass of $A$ belongs to $f $.
(2) The union of any nest in $f$ (if $g$ is a subclass of $f$ and $g$ is a nest, then $g$ is a nest in $f$) is the member of $f$. Here is the Coq formal statement of them:

\begin{lstlisting}
Proposition Property_FinChar : $\forall$ f: Class,
  Finite_Char f /\ f $\neq$ $\emptyset$ -> ($\forall$ A B, A $\in$ f /\ B $\subset$ A -> B $\in$ f) /\
  ($\forall$ g, g $\subset$ f /\ Nest g -> ($\bigcup$g) $\in$ f).
\end{lstlisting}

\begin{proof}
First of all, we prove the first property. Based on the definition of the set of finite character, any finite subclass of $B$ is also a finite subclass of $A$. Therefore, $B$ is a member of $f$. Secondly, we prove the second property. Assume that $D$ is a finite subclass of $\bigcup g$. There are finite members $C_{1}$, $C_{2}$, $\cdots$, $C_{n}$ in $g$ so that $D\subset C_{1} \cup C_{2} \cup \cdots \cup C_{n}$ is established. Since $g$ is a nest, $D$ is a member of $g$. Then $\bigcup g$ is a member of $f$.
\end{proof}

The following are some basic definitions needed to prove Zorn's lemma and the well-ordering theorem, including partial orders, total orders, well orders, chains, initial segment and so on.

In the order theory of mathematics, a partially ordered set generalizes the intuitive concept of an ordering of the elements of a set. Partial orders can generalize total orders and well orders can be obtained from total orders. The definitions of partial orders and partially ordered set are as follows.

\begin{defi}[Partial Order, Partially Ordered Set]
A (non-strict) partial order is a binary relation $\leq$ over a set $X$ (a set with a partial order is called a partially ordered set) satisfying particular properties which are discussed below. That is, for all $a$, $b$, and $c$ in $X$, it must satisfy:

\par\setlength\parindent{0.5em}
(1) $a \leq a$ (reflexivity: every element is related to itself).

(2) If $a \leq b$ and $b \leq a$, then $a = b$ (antisymmetry: two distinct elements cannot be related in both directions).

(3) If $a \leq b$ and $b \leq c$, then $a \leq c$ (transitivity: if a first element is related to a second element, and, in turn, that element is related to a third element, then the first element is related to the third element).
\end{defi}

We formally define reflexivity, antisymmetry and transitivity first. They can be easily translated in Coq:

\begin{lstlisting}
Definition Reflexivity le X : Prop := $\forall$ a, a $\in$ X -> Rrelation a le a.
Definition Antisymmetry le X : Prop := $\forall$ a b, a $\in$ X /\ b $\in$ X ->
  (Rrelation a le b /\ Rrelation b le a -> a = b).
Definition Transitivity le X : Prop := $\forall$ a b c, (a$\in$X /\ b$\in$X /\ c$\in$X) /\
  (Rrelation a le b /\ Rrelation b le c) -> Rrelation a le c.
\end{lstlisting}

The formalization of partial orders is on the basis of the above three properties. The binary relation $\leq$ over a set $X$ is formalized as a subclass of the Cartesian product $X\times X$. Because of a partially ordered set is a set with a partial order, we formalize it directly through partial sets.

\begin{lstlisting}
Definition PartialOrder le X : Prop :=
  Ensemble X /\ (Relation le /\ le $\subset$ (X $\times$ X)) /\
  Reflexivity le X /\ Antisymmetry le X /\ Transitivity le X.
Definition PartialOrderSet X le := PartialOrder le X.
\end{lstlisting}

On the basis of partial orders, we define the upper (lower) bound and the maximal (minimal) element.

\begin{defi}[Upper (Lower) Bound]
Assume that $\leq$ is a partial order on the non-empty set $X$. For a subclass $A$ of $X$, an element $x$ in $X$ is an upper bound of $A$ if $a\leq x$, for each element $a$ in $A$. Similarly, an element $x$ in $X$ is a lower bound of $A$ if $a\leq x$, for each element $a$ in $A$.
\end{defi}

\begin{lstlisting}
Definition BoundU x A X le : Prop := PartialOrder le X /\ X $\neq$ $\emptyset$ ->
  x $\in$ X /\ A $\subset$ X /\ ($\forall$ a, a $\in$ A -> Rrelation a le x).
Definition BoundL x A X le : Prop := PartialOrder le X /\ X $\neq$ $\emptyset$ ->
  x $\in$ X /\ A $\subset$ X /\ ($\forall$ a, a $\in$ A -> Rrelation x le a).
\end{lstlisting}

If a partial order is a relation over a set $X$, it is quite possible to find a maximal (minimal) element of the subset of $X$. So when we define a maximal (minimal) element, we do not specify a specific set for the binary relation.

\begin{defi}[Maximal (Minimal) Element]
Assume that $X$ is not empty and $\leq$ is an ordering of $X$. $x$ is a maximal (minimal) element of $X$ if for all $y\in X$, $x\leq y$ and $x\neq y$ ($y\leq x$ and $y\neq x$) are false.
\end{defi}

\begin{lstlisting}
Definition MaxElement x X le : Prop :=
  X $\neq$ $\emptyset$ -> x $\in$ X /\ ($\forall$ y, y $\in$ X -> $\sim$ (Rrelation x le y /\ x $\neq$ y)).
Definition MinElement x X le : Prop :=
  X $\neq$ $\emptyset$ -> x $\in$ X /\ ($\forall$ y, y $\in$ X -> $\sim$ (Rrelation y le x /\ y $\neq$ x)).
\end{lstlisting}

Next we define total orders. If a partial order is also a connex relation, then it is a total order. A relation having the connex property means that any pair of elements in the set of the relation are comparable under the relation. The specific definitions and the Coq implementations of them are as follows.

\begin{defi}[Total Order, Totally Ordered Set]
Assume that $\leq$ is a partial order on a set $X$. $\leq$ is a total order on $X$ and $(X,\leq)$ is a totally ordered set iff $a\leq b$ or $b\leq a$ for every $a,b\in X$ (connex property).
\end{defi}

\begin{lstlisting}
Definition TotalOrder le X := PartialOrder le X /\
  ($\forall$ x y, x $\in$ X /\ y $\in$ X -> Rrelation x le y \/ Rrelation y le x).
Definition TotalOrderSet X le := TotalOrder le X.
\end{lstlisting}

With the definition of total orders, we can define chains. The definition of chains will be used in Zorn's lemma.

\begin{defi}[Chain]
Assume that $\leq$ is a partial order on the set $X$. If $A$ is a non-empty subclass of $X$ and the relation $\leq \cap\, (A\times A)$ is a total order on $A$, then $A$ is a chain of $X$.
\end{defi}

\begin{lstlisting}
Definition Chain A X le : Prop :=
  PartialOrder le X -> (A $\subset$ X /\ A $\neq$ $\emptyset$) /\ TotalOrder (le $\cap$ (A $\times$ A)) A.
\end{lstlisting}

A well order is a relation over a set here and it is different from the definition in the ``axiomatic set theory'' formal system. We can define it by adding a property to a total order.

\begin{defi}[Well Order, Well-Ordered Set]
Assume that $\leq$ is a total order on the set $X$. $\leq$ is a well order on $X$ and $(X,\leq)$ is a well-ordered set iff every non-empty subclass of $X$ has a least element.
\end{defi}

According to the definition of total orders, every element in the set of the relation are comparable under the relation. Therefore, the minimal element in well orders is equivalent to the least element. We formalize well orders through the formalization of minimal elements.

\begin{lstlisting}
Definition WellOrder le X : Prop :=
  TotalOrder le X /\ ($\forall$ A, A $\subset$ X /\ A $\neq$ $\emptyset$ -> $\exists$ z, MinElement z A le).
Definition WellOrderSet X le := WellOrder le X.
\end{lstlisting}

Then we define the initial segment based on well orders. The definition of the initial segment will be involved in the well-ordering theorem.

\begin{defi}[Initial Segment]
Assume that $(X,\leq)$ is a well-ordered set. $Y\subset X$ is an initial segment of $X$ iff it satisfies the condition: if $u\leq v$ for $u\in X$ and $v\in Y$, then $u\in Y$.
\end{defi}

\begin{lstlisting}
Definition Initial_Segment Y X le : Prop := Y $\subset$ X /\ WellOrderSet X le /\
  ($\forall$ u v, (u $\in$ X /\ v $\in$ Y /\ Rrelation u le v) -> u $\in$ Y).
\end{lstlisting}

On the basis of these basic definitions and properties, we complete the cyclic proof of equivalence between AC and its related theorems in the following section.

\section{Formal Proof of the Equivlance}
\label{sec:4}
In this section, we present the formal proof of AC and its equivalent theorems. As shown in Figure\,1, we start from AC to prove Tukey's lemma, the Hausdorff maximal principle, the maximal principle, Zermelo's postulate, Zorn's lemma and the well-ordering theorem in turn. We prove AC through Zermelo's postulate and the well-ordering theorem finally, thus completing the cyclic proof of the equivalence between AC and these theorems. Before each theorem is proved, we will give its formal description.

\begin{figure}[ht]
  \centering
  \includegraphics[height=6cm,width=12.5cm]{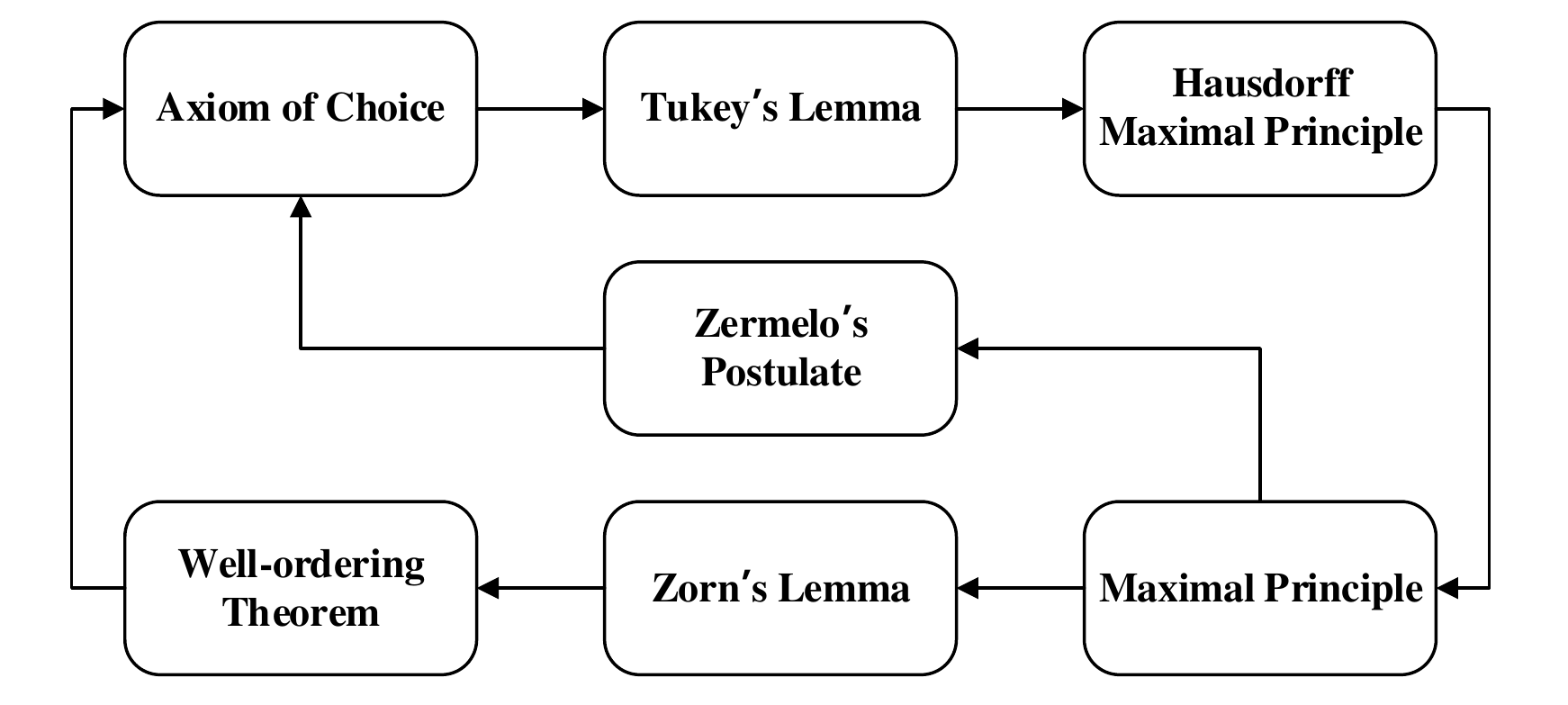}
  \caption{The relation of AC and its equivalent theorems}
\end{figure}

\subsection{Tukey's Lemma}
First, we regard AC as an axiom and present the content of it according to the definition of sets and the choice function. We give a Coq formal description of it at the same time.

\vskip 0.3cm
{\noindent{\bf Axiom of Choice}\quad{\itshape Every set has a choice function.}}

\begin{lstlisting}
Axiom Choice_Axiom : $\forall$ X, Ensemble X -> $\exists$ c, Choice_Function c X.
\end{lstlisting}

The axiom AC will be used in the proof process of Tukey's lemma. Tukey's lemma was formulated independently by Oswald Teichm\"{u}ller in 1939 and John Tukey in 1940 \cite{J73}. Tukey's lemma states the following:

\begin{thm}[Tukey's Lemma]
There is a maximal member of each non-empty set of finite character.
\end{thm}

We formalize the theorem based on Definition\,3.2 and Definition\,3.4. In our formalization, it is expressed as follows:

\begin{lstlisting}
Theorem Tukey : $\forall$ f, Finite_Character f /\ f $\neq$ $\emptyset$ -> $\exists$ x, MaxMember x f.
\end{lstlisting}

Then some concepts and formulas which involved in the proof processes are formalized before beginning to prove the theorem. Assume that $f$ is a non-empty set of finite character, thus $\bigcup f$ is a set. According to AC, there is a choice function $c$ of the set $\bigcup f$. For any member $F$ of the nonempty set $f$, we construct a larger set $\{ x : (x\in\bigcup f) \wedge\,(F\cup \{x\})\in f\}$ and record it as $En\_F'$. We formally define it as follows:

\begin{lstlisting}
Definition En_F' F f : Class := \{ fun x => x $\in$ ($\bigcup$f) /\ (F $\cup$ [x]) $\in$ f \}.
\end{lstlisting}

We introduce the correctly usage method of the $Classifier$ here.
As mentioned in the preceding section\,2, the input of the $Classifier$ is an item with type $Class\rightarrow Prop$. The first parameter $Class$ represents any variable of the classifier and the second parameter $Prop$ means the expression that the variable satisfies.
The $\lambda$-abstraction meets our requirements above and it is implemented through function $\texttt{fun}$ in Coq.

According to the definition of $En\_F'$, we can get a conclusion that either $En\_F'$ contains $F$ or $En\_F'$ equals to $F$. Thus we define a function $\chi$ according to it. The domain of $\chi$ is $f$ and the range of $\chi$ is a subclass of $f$. This function is described in detail as follows:

for any $F\in f$,
\begin{equation}
\chi(F)=\left\{
\begin{array}{ll}
F\cup \{c(F^{\prime }-F)\}, &\qquad F^{\prime }-F\neq \phi  \\
F, &\qquad F^{\prime }-F=\phi
\end{array}%
\right.
\end{equation}

For the Coq formalization of the function $\chi$, we divide it into three parts. The first part defines a decidable function $eq\_dec$ which can be used for any $A$ whose type is Type. In order to implement it, the $sumbool$ function in the Coq library is used. $sumbool$ is a Boolean type equipped with the justification of their value.

\begin{lstlisting}
Inductive sumbool (P1 P2: Prop) : Set :=
  | left : P1 -> { P1 } + { P2 }
  | right : P2 -> { P1 } + { P2 }
\end{lstlisting}

We can get $P1$ and $P2$ directly through two constructors left and right. Formally the function $eq\_dec$ is defined as follows:

\begin{lstlisting}
Definition eq_dec (A: Type) := forall x y: A, { x = y } + { x $\neq$ y }.
\end{lstlisting}

In the second part, we give $eq\_dec$ a specific type $Class$ as follows.

\begin{lstlisting}
Parameter beq : eq_dec Class.
\end{lstlisting}

In the third part, on the basis of the function $beq$, we define the function $Fun\_X$ by using pattern matching in Coq. It is straightforward to implement in Coq:

\begin{lstlisting}
Definition Fun_X (F f c: Class) : Class :=
  match beq ((En_F' F f) $\sim$ F) $\emptyset$ with
  | left _ => F
  | right _ => F $\cup$ [c[(En_F' F f) $\sim$ F]]
  end.
\end{lstlisting}

Through the definition of the function $\chi$, we can convert the proof target into proving whether there is a member $F$ of $f$, so that $\chi(F) = F$. Then we define the t-Subclass based on the function $\chi$.

\begin{defi}[t-Subclass]
$g$ is a t-Subclass of $f$ iff $g$ is a subclass of $f$ and it meets the following conditions:

\vskip 0.1cm
\par\setlength\parindent{0.5em} (1) $\emptyset$ is a member of $g$;

\vskip 0.1cm
\par\setlength\parindent{0.5em} (2) If $F$ is a member of $g$, then $\chi(F)\in g$;

\vskip 0.1cm
\par\setlength\parindent{0.5em}(3) If $\mathcal{L}$ is a nest in $g$ $(\mathcal{L}\subset A$ and $\mathcal{L}$ is a nest$)$, then $\bigcup\mathcal{L}\in g$.
\end{defi}

Its statement in Coq is the following one:

\begin{lstlisting}
Definition tSubclass (g f c: Class) : Prop :=
  g $\subset$ f /\ $\emptyset$ $\in$ g /\ ($\forall$ F, F $\in$ g -> (Fun_X F f c) $\in$ g) /\
  ($\forall$ L, L $\subset$ g /\ Nest L -> ($\bigcup$L) $\in$ g).
\end{lstlisting}

Next we introduce some equations. Let $f'0$ be the intersection of all the t-Subclass of the set $f$, as shown in equation\,(4.2).
\begin{equation}
  f'0 = \bigcap\,\{ g : tSubclass\,g\,f\,c \}
\end{equation}

If we can prove that $f'0$ is a nest, then we prove the theorem. The specific proof process will be given in section\,4.1.4. In order to prove $f'0$ is a nest, we construct $\mu$, $f'1$ and $\upsilon$ as shown in the following equations.

For every member $C$ of $f'0$, let
\begin{equation}
  \mu(C) = \{ A : A\in f'0\,\wedge\,(A\subset C\vee C\subset A) \}
\end{equation}
and
\begin{equation}
  f'1 = \{ C : C\in f'0\,\wedge\,\mu(C)=f'0\}
\end{equation}

For every member $D$ of $f'1$, let
\begin{equation}
  \upsilon(D) = \{ A : A\in f'0\,\wedge\,(A\subset D \vee \chi(D)\subset A) \}
\end{equation}

These equations can be easily formalized in Coq:

\begin{lstlisting}
Definition En_f'0 f c := $\bigcap$ \{ fun g => tSubclass g f c \}.
Definition En_u C f c : Class :=
  \{fun A => A $\in$ (En_f'0 f c) /\ (A $\subset$ C \/ C $\subset$ A)\}.
Definition En_f'1 f c : Class :=
  \{ fun C => C $\in$ (En_f'0 f c) /\ (En_u C f c) = (En_f'0 f c) \}.
Definition En_v D f c : Class :=
  \{ fun A => A $\in$ (En_f'0 f c) /\ (A $\subset$ D \/ (Fun_x D f c) $\subset$ A) \}.
\end{lstlisting}

After some concepts and equations are built, we prove two properties. They will be often used in the later proof and they have been proved in Coq. The first property is about the function $\chi$. It states that every member of $f$ is a subclass of the function $\chi(F)$.

\begin{lstlisting}
Proposition Property_x : $\forall$ (c F f: Class),
  Choice_Function c ($\bigcup$ f)-> F $\in$ f -> F $\subset$ (Fun_X F f c).
\end{lstlisting}

\begin{proof}
We introduce an assumption that either $F'=F$ or $F'\neq F$. It is the law of excluded middle. Then the proposition can be proved by the definition of the function $\chi$.
\end{proof}

The second property is about $f'0$ which is defined in equation\,(4.2). The class $f'0$ is the least t-Subclass of $f$, that is $f'0$ is a t-Subclass of $f$ and $f'0\subset g$ if $g$ is a t-Subclass of $f$.

\begin{lstlisting}
Proposition Property_f'0 : $\forall$ (f c: Class),
  Finite_Character f /\ f $\neq$ $\emptyset$ -> Choice_Function c ($\bigcup$ f) ->
  tSubclass (En_f'0 f c) f c /\ ($\forall$ g, g $\subset$ f /\ tSubclass g f c ->
  (En_f'0 f c) $\subset$ g).
\end{lstlisting}

\begin{proof}
We can prove that $f$ is a t-Subclass of $f$ according to $f$ is a nonempty set of finite character. It is easy to verify that the intersection of any t-Subclass is still a t-Subclass by the conclusion $f$ is a t-Subclass. Thus $f'0$ is a t-Subclass. Finally, on the basis of equation\,(4.2), we can prove that $f'0$ is the least t-Subclass.
\end{proof}

The following will be five steps to complete the proof of Tukey's lemma. In the first three steps, we prove $f'0$ is a nest. We prove that $\chi(F) = F$ if $F=\bigcup f'0$ in the fourth step. Finally, we directly prove Tukey's lemma based on the above conclusion.

\subsubsection{Step I}

The first step is to prove that if $D$ is a member of $f'1$, then $\upsilon(D)$ is a t-Subclass of the non-empty set of finite character $f$.

In the formalization, it should be noted that we must declare the parameter $c$ when using the t-Subclass. So we have to add the condition that $c$ is a choice function of the set $\bigcup f$ to the lemma. The formalization is as follows:

\begin{lstlisting}
Lemma LemmaT1 : $\forall$ (f c: Class),
  Finite_Character f /\ f $\neq$ $\emptyset$ -> Choice_Function c ($\bigcup$ f) ->
  ($\forall$ D, D $\in$ (En_f'1 f c) -> tSubclass (En_v D f c) f c).
\end{lstlisting}

\begin{proof}
First of all, we have $\upsilon(D)\subset f$ by the definition of $\upsilon(D)$ in equation\,(4.5) and $Property\_f'0$. So $A$ is a member of $f$ can be obtained for any $A\in\upsilon(D)$. This conclusion will be used repeatedly in subsequent proofs. We can prove $f'0$ is a t-Subclass based on $Property\_f'0$. Then we gradually prove that $\upsilon(D)$ satisfies the three conditions of t-Subclass as follows:

\vskip 0.1cm
\par\setlength\parindent{0.5em} 1. Obviously, $\emptyset$ is a member of $\upsilon(D)$ according to equation\,(4.5).

\vskip 0.1cm
\par\setlength\parindent{0.5em} 2. Let's verify that: If $A$ is a member of $\upsilon(D)$, then $\chi(A)$ is a member of $\upsilon(D)$. Then we will discuss in three cases next:

\vskip 0.1cm
\par\setlength\parindent{1em} 2.1. $A\subset D$ and $A\neq D$. We prove there must be $\chi(A)\subset D$ at this time. Here we prove it by reduction to absurdity, we firstly assume that $\chi(A)\subset D$ is false. $D$ is a member of $f'0$ and $\mu(D)=f'0$ because of $D$ is a member of $f'1$. Then we have $A\in f'0$ according to $A\in\upsilon(D)$. Since $f'0$ is a t-Subclass from $Property\_f'0$, $\chi(A)$ is a member of $f'0$. After combining the above conclusions, we have $\chi(A)$ belongs to $\mu(D)$, so $\chi(A)\subset D$ or $D\subset\chi(A)$. According to our hypothesis that $\chi(A)\subset D$ is false, we can prove $D\subsetneq\chi(A)$. Therefore, $\chi(A)$ is at least two points more than $A$. It is obvious that $\chi(A)$ is equal to $A$ or $\chi(A)$ is one point more than $A$ on the basis of the definition of function $\chi$ in equation\,(4.1). So there is a contradiction here and we have $\chi(A)\subset D$. To sum up, $\chi(A)$ is a member of $\upsilon(D)$.

\vskip 0.1cm
\par\setlength\parindent{1em} 2.2. $A=D$. We can get a conclusion that $\chi(A)=\chi(D)$ when $A=D$. Since $\chi(D)\in\upsilon(D)$ is obvious, $\chi(A)$ is a member of $\upsilon(D)$.

\vskip 0.1cm
\par\setlength\parindent{1em} 2.3. $\chi(D)\subset A$. We obtain that $A\subset\chi(A)$ from $Property\_x$. Moreover, $\chi(D)\subset\chi(A)$ can be proved easily. Thus $\chi(A)$ is a member of $\upsilon(D)$.

\vskip 0.1cm
\par\setlength\parindent{0.5em} 3. If $\psi$ is a nest in $\upsilon(D)$, we will discuss in two cases because the definition of $\upsilon(D)$. If all members of $\psi$ are contained in $D$, then $\bigcup\psi\subset D$; If $\psi$ has one member that contains $\chi(D)$, then $\chi(D)\subset\bigcup\psi$. And because of $f'0$ is a t-Subclass, $\bigcup\psi$ is a member of $f'0$. We prove that $\bigcup\psi\in\upsilon(D)$ based on the definition of $\upsilon(D)$. So far, we have completed the proof of $LemmaT1$.
\end{proof}

\subsubsection{Step II}

$LemmaT1$ has proved that $\upsilon(D)$ is a t-Subclass for every member $D$ of $f'1$. In the second step, we prove that $f'1$ satisfies the second condition of t-Subclass, that is if $D$ is a member of $f'1$, then $\chi(D)$ is a member of $f'1$ as well.

\begin{lstlisting}
Lemma LemmaT2 : $\forall$ (f c: Class),
  Finite_Character f /\ f $\neq$ $\emptyset$ -> Choice_Function c ($\bigcup$ f) ->
  ($\forall$ D, D $\in$ (En_f'1 f c)-> (Fun_X D f c) $\in$ (En_f'1 f c)).
\end{lstlisting}

\begin{proof}
According to equation\,(4.3) and equation\,(4.5), it is obvious that $\upsilon(D)\subset \mu(\chi(D))$ for every $D\in f'1$. It can be known that $\upsilon(D)$ is a t-Subclass from $LemmaT1$ and $f'0$ is the least t-Subclass from $Property\_f'0$. In that way, $f'0\subset\upsilon(D)$. Therefore, we can get a conclusion that $f'0\subset\mu(\chi(D))$. Based on equations above, $\mu(\chi(D))\subset f'0$ can be proved easily. Thus $\mu(\chi(D))=f'0$, namely, $\chi(D)$ is a member of $f'1$.
\end{proof}

\subsubsection{Step III}
Next we prove $f'0$ is a nest. It can be proved by $LemmaT2$. The specific description and Coq statement of the lemma is as follows:

\begin{lstlisting}
Lemma LemmaT3 : $\forall$ f c, Finite_Character f /\ f $\neq$ $\emptyset$ ->
  Choice_Function c ($\bigcup$ f) -> Nest (tSub_f'0 f c).
\end{lstlisting}

\begin{proof}
According to the definition of $f'1$ in equation\,(4.4), it is obvious that $f'1$ is a nest in $f'0$. If we prove that $f'1$ is a t-Subclass, then we prove that $f'0$ is a nest. It can be known that $f'0\subset f'1$ because $f'0$ is the least t-Subclass of $f$. Then we get the conclusion that $f'0=f'1$ according to $f'0\subset f'1$ and $f'1\subset f'0$. Thus $f'0$ is a nest and $LemmaT3$ is proved.

Next, we prove $f'1$ is a t-Subclass. Since $f'1$ is contained in $f'0$ and $f'0$ is contained in $f$, $f'1$ is contained in $f$. It must meet three conditions of t-Subclass:

\vskip 0.1cm
\par\setlength\parindent{0.5em} (1) It is obvious that $\emptyset$ is a member of $f'1$;

\vskip 0.1cm
\par\setlength\parindent{0.5em} (2) It can be known that $\chi(D)\in f'1$ for every $D\in f'1$ from $LemmaT2$.

\vskip 0.1cm
\par\setlength\parindent{0.5em} (3) Assume that $\omega$ is a nest in $f'1$, $\mu(C)=f'0$ for any member $C$ of $\omega$. According to the definition of $f'0$, $A\subset C$ or $C\subset A$ for every member $A$ of $f'0$. Here we discuss it in two cases: First, if all members of $\omega$ are contained in $A$, then $\bigcup\omega\subset A$ can be proved. Second, if $\omega$ has one member that includes $A$, then $A\subset\bigcup\omega$ can be proved. That is to say $\mu(\bigcup\omega)=f'0$. Accordingly, $\bigcup\omega$ is a member of $f'1$. Overall, the proof of $LemmaT3$ is completed.
\end{proof}

\subsubsection{Step IV}
After proving that $f'0$ is a nest, we can prove that $\bigcup f'0$ is a member of $f$ and $\chi(\,\bigcup f'0\,) = \bigcup f'0$.

\begin{lstlisting}
Lemma LemmaT4 : $\forall$ f c, Finite_Character f /\ f $\neq$ $\emptyset$ ->
  Choice_Function c ($\bigcup$ f) -> ($\bigcup$ (tSub_f'0 f c)) $\in$ f /\
  (Function_x ($\bigcup$ (tSub_f'0 f c)) f c) = $\bigcup$ (tSub_f'0 f c).
\end{lstlisting}

\begin{proof}
Let $F=\bigcup f'0$, it can be known that $F\subset\chi(F)$ from $Property\_x$ which has been proved above. $f'0$ is a t-Subclass based on $Property\_f'0$. $f'0\subset f'0$ is obvious and $f'0$ is a nest according to $LemmaT3$. Therefore, $F$ is a member of $f'0$. Then we prove $\chi(\bigcup f'0) = \bigcup f'0$. The conclusion that $\chi(F)$ belongs to $f'0$ can be obtained based on Definition\,4.2. Thus it is easy to prove that $\chi(F)$ is contained in $F$, then we can get $\chi(F)=F$. The proof of $LemmaT4$ is completed.
\end{proof}

\subsubsection{Step V}
In the last step, Tukey's lemma can be proved by $LemmaT4$. We need to prove that a nonempty set of finite character must has a maximal member. The specific process of proof is as follows:

\begin{proof}
According to previous hypotheses, $f$ is a nonempty set of finite character and there is a choice function $c$ of the set $\bigcup f$ according to AC. The domain of function $c$ is $2^{\bigcup f}-\{\emptyset\}$, the range of function $c$ is a subclass of $\bigcup f$ and $c(A)$ is a member of $A$ for every $A\in dom(c)$. On the basis of $LemmaT4$, there is a member $F=\bigcup f'0$ of $f$ and $\chi(F) = F$. Thus $F=F'$ according to the definition of the function $\chi$. Therefore, the class $F$ is a maximal member of $f$. In conclusion, Tukey's lemma is fully proved.
\end{proof}

\subsection{Hausdorff Maximal Principle}
The Hausdorff maximal principle can be proved according to Tukey's lemma. A specific description of the Hausdorff maximal principle is given below.

\begin{thm}[Hausdorff Maximal Principle]
If $A$ is a set and $N$ is a nest in $A$, then there is a maximal nest in $A$ which contains $N$.
\end{thm}

We construct a set $\{ n : n\subset A\wedge Nest\,n\}$ whose member is the nest in $A$. Then the maximal nest is obtained by finding the maximal member of the set. The formal description of the theorem in Coq is as follows:

\begin{lstlisting}
Theorem Hausdorff : $\forall$ A N, Ensemble A -> N $\subset$ A /\ Nest N ->
  ($\exists$ u, MaxMember u \{ fun n => n $\subset$ A /\ Nest n \} /\ N $\subset$ u).
\end{lstlisting}

The proof of the theorem based on two lemmas, so we prove them first. The description of the first lemma is that if $f$ is a set of finite character and $A$ is a member of $f$, there is a maximal member of $f$ and it contains $A$. Its statement in Coq is as follows:

\begin{lstlisting}
Lemma LemmaH1 : $\forall$ (f A: Class),
  Finite_Character f -> A $\in$ f -> ($\exists$ M, MaxMember M f /\ A $\subset$ M).
\end{lstlisting}

\begin{proof}
Let $\mathcal{F}_{1} = \{ F : F\in f \wedge (F\cup A)\in f \}$ and $\mathcal{F}_{1}$ is not an empty set because of $A\in \mathcal{F}_{1}$. The formalization of it is as follows:

\begin{lstlisting}
Definition En_f1 f A := \{ fun F => F $\in$ f /\ (F $\cup$ A) $\in$ f \}.
\end{lstlisting}

We first prove that $\mathcal{F}_{1}$ is a set of finite character. Assume that $B$ is a member of $\mathcal{F}_{1}$. If $B_{1}$ is a finite subclass of $B$, then $B_{1}$ is a member of $f$ according to Definition\,3.4. We can prove that $B_{1}\cup A\in f$ on the basis of $B_{1}\cup A\subset B\cup A$. Conversely, if every finite subclass of $C\in f$ is belongs to $\mathcal{F}_{1}$, we consider whether $C\in\mathcal{F}_{1}$ is established. If $A_{1}$ is a finite subclass of $C\cup A$, let $C_{1} = C\cap A_{1}$. We can prove $C_{1}$ is a finite subclass of $C$ based on $Property\_FinChar$ in section\,3. Consequently, $C_{1}$ is a member of $\mathcal{F}_{1}$ and $C_{1}\cup A$ is a member of $f$. Then $A_{1}\subset (C_{1}\cup A)$ can be proved, and $A_{1}\in f$ is established. Therefore, $C\in\mathcal{F}_{1}$ is established and we have proved that $\mathcal{F}_{1}$ is a set of finite character.

Next we prove that there is a maximal member of $f$ and it contains $A$. We have proved that $\mathcal{F}_{1}$ is a set of finite character above. Thus $\mathcal{F}_{1}$ has a maximal member $M$ on the basis of Tukey's lemma, and $M\cup A\in \mathcal{F}_{1}$. Since $M$ is the maximal member of $\mathcal{F}_{1}$, $M = M\cup A$ and $A\subset M$ can be proved. Finally, we prove that $M$ is a maximal member of $f$. If there is a member $K$ of $f$ and $M\subset K$, then we can prove $K\cup A = K$ because of $A\subset M$. Therefore, $K\in \mathcal{F}_{1}$ is proved by $K\in f$ and $K\cup A = K$. Because of $M$ is the maximal member of $\mathcal{F}_{1}$ and $M\subset K$, we can get $M = K$. Thus $M$ is a maximal member of $f$.
\end{proof}

Then we prove the second lemma. If $A$ is a set, then the set which is consist of all nests in $A$ is of finite character. In formalization, we construct a set $\{ n : n\subset A \wedge Nest\, n\}$ to represent the set which is consist of all nests in $A$ as follows:

\begin{lstlisting}
Lemma LemmaH2 : $\forall$ (A: Class),
  Ensemble A -> Finite_Character \{ fun n => n $\subset$ A /\ Nest n \}.
\end{lstlisting}

\begin{proof}
First of all, according to the Definition\,3.4, the target of proof can be changed as follows,

\vskip 0.1cm
\par\setlength\parindent{0.5em} (1) If $N\in\{n : n\subset A\wedge Nest\,n\}$ and $z$ is a finite subclass of $N$, then $z$ is a member of $\{n : n\subset A\wedge Nest\,n\}$;

\vskip 0.1cm
(2) If every finite subclass of the set $N$ belongs to $\{n : n\subset A\wedge Nest\,n\}$, then $N$ is a member of $\{n : n\subset A\wedge Nest\,n\}$.

\vskip 0.1cm
\par\setlength\parindent{1.5em}We prove the first part first. We apply the classification axiom-scheme to the proposition, and the condition of proof becomes $N\subset A$, $N$ is a nest and $z$ is a finite subclass of $N$, the target of proof becomes $z\subset A$ and $z$ is a nest. $z\subset A$ can be proved by the condition that $z\subset N$ and $N\subset A$. Then $z$ is a nest can be proved according to Definition\,3.3.

Next we prove the second part. Similar to the above proof process, we can also convert the proof goal into $N \subset A$ and $N$ is a nest. If $x$ is a member of $N$, then $\{x\}\subset N$ and $\{x\}$ is finite can be proved. Thus $\{x\}$ belongs to $\{n : n\subset A\wedge Nest\,n\}$, this also means that $\{x\}\subset A$ is established. Therefore, $N\subset A$ is proved. Assume that $y$ and $z$ are members of $N$. We prove the unordered pair $\{x\,y\}$ is finite and $\{x\,y\}\subset N$. According to the condition, $\{x\,y\}\subset A$ and $\{x\,y\}$ is a nest. Either $x\subset y$ or $y\subset x$ can be proved by the definition of nest. Consequently, $N$ is a nest.
\end{proof}

Finally, the Hausdorff maximal principle can be easily proved on the basis of $LemmaH1$ and $LemmaH2$. The specific proof process is as follows:

\begin{proof}
According to the conditions of Theorem\,4.3, $A$ is a set and $N$ is a nest in $A$. The $\{n : n\subset A \wedge Nest\, n\}$ is a set of finite character can be proved on the basis of $LemmaH2$. Then we apply $LemmaH1$ on the set, there is a maximal member of it which contains $N$. In summary, the Hausdorff maximal principle is proved.
\end{proof}

The above process is verified in Coq in only 4 lines. As shown in Figure 2, the proof process is highly readable \footnote{The theorem\,33 is a theorem about the existence of sets in Kelley's set theory \cite{K55}.}.

\begin{figure}[ht]
  \centering
  \includegraphics[height=6cm,width=11cm]{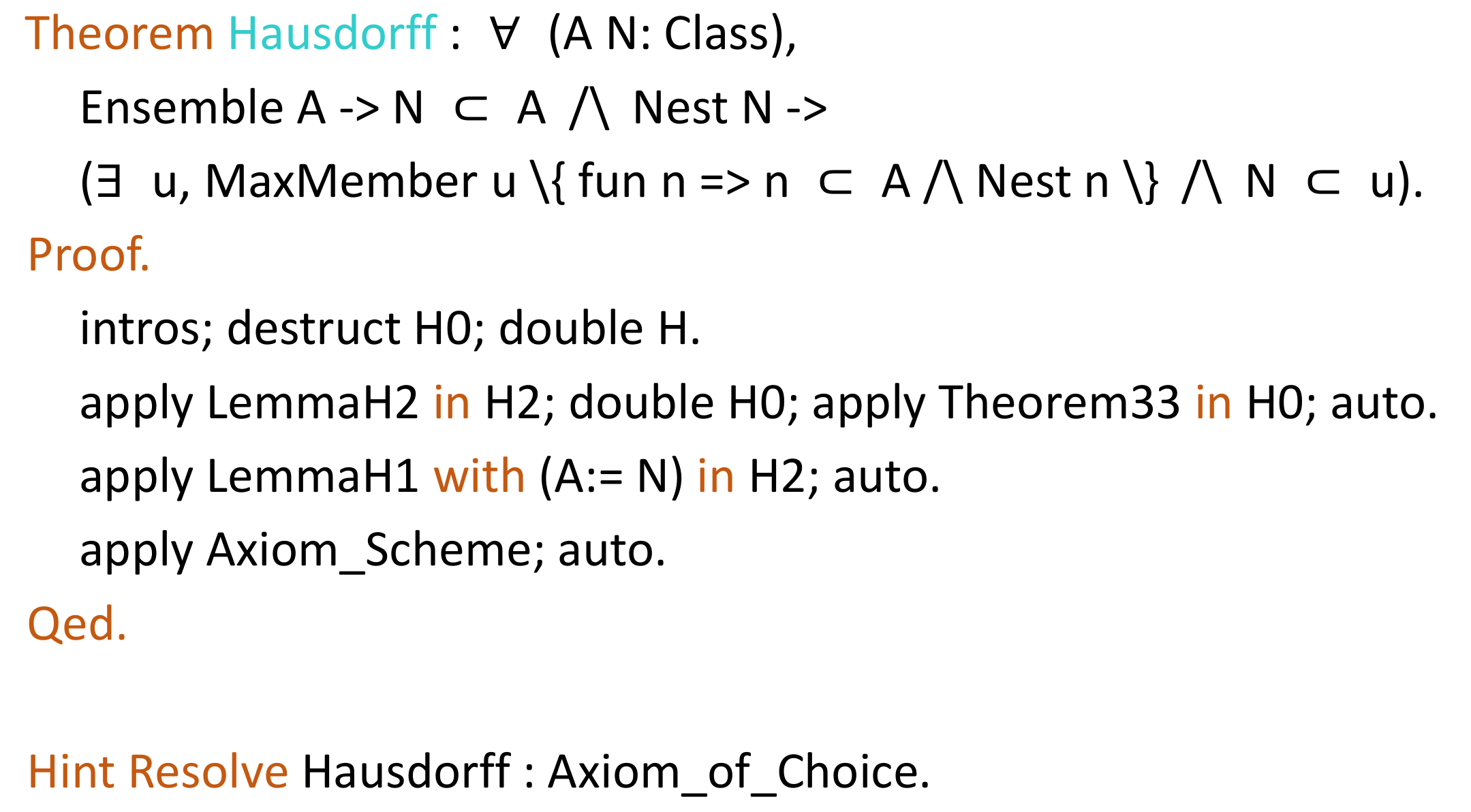}
  \caption{The Proof of the Hausdorff maximal principle in Coq}
\end{figure}

\subsection{Maximal Principle}
The maximal principle can be proved according to the Hausdorff maximal principle. The description of the maximal principle is as follows.

\begin{thm}[Maximal Principle]
There is a maximal member of a set $A$, provided that for each nest in $A$ there is a member of $A$ which contains every member of the nest.
\end{thm}

It is straightforward to formalize the theorem in Coq as follows:

\begin{lstlisting}
Theorem MaxPrinciple : $\forall$ (A: Class),
  Ensemble A -> ($\forall$ n, n $\subset$ A /\ Nest n -> $\exists$ N, N $\in$ A /\
  ($\forall$ u, u $\in$ n -> u $\subset$ N)) -> $\exists$ M, MaxMember M A.
\end{lstlisting}

\begin{proof}
Assume that $n$ is a nest in the set $A$. We can get a conclusion that $u$ is a maximal nest in $A$ and $u$ contains $n$ according to the Hausdorff maximal principle. In our formalization, we construct the set $\{ n : n\subset A \wedge\, Nest\,n \}$. The maximal nest $u$ is obtained by taking the maximal member from this set. Next there is a member $N$ of $A$ which contains every member of the nest $u$ on the basis of the condition of the theorem. If there is a member $K$ of $A$ and $N\subsetneq K$, we can prove $u\cup \{K\}$ is nest which contains $u$. Since $u$ is the maximal nest, $u\cup \{K\}$ and $u$ are equal. That means $K\in u$. Therefore, the conclusion of $N=K$ is established. To summarize, $N$ is a maximal member of the set $A$. The maximal principle is completely proved.
\end{proof}

\subsection{Zermelo's Postulate}
We can prove Zermelo's postulate according to the maximal principle. The description of Zermelo's postulate is as follows:

\begin{thm}[Zermelo's Postulate]
If $A$ is a disjoint family of non-empty sets, then there is a set $C$ such that $D \cap C$ is a singleton set for every $D$ in $A$.
\end{thm}

As follows, we describe the disjoint family of non-empty sets by two properties.

\begin{lstlisting}
Theorem Zermelo : $\forall$ A: Class,
  Ensemble A -> $\emptyset$ $\notin$ A -> ($\forall$ x y, x $\in$ A /\ y $\in$ A -> x $\neq$ y -> x $\cap$ y = $\emptyset$) ->
  ($\exists$ C, Ensemble C /\ ($\forall$ D, D $\in$ A -> $\exists$ x, D $\cap$ C = [x])).
\end{lstlisting}

\begin{proof}
If $A$ is a disjoint family of non-empty sets, then let $X = \bigcup A$.
For every non-empty subset $B$ of $A$, we construct a set $T_{B}$. $T_{B}$ is consist of the subset $K$ of $X$, $K$ satisfies the following conditions: (1) For every $D\in A \sim B$, $D \cap K = \emptyset$; (2) For every $D\in B$, $D \cap K$ is a singleton set. The Coq statement of the set $T_{B}$ is as follows:

\begin{lstlisting}
Definition En_TB B A : Class :=
  \{ fun K => K $\subset$ ($\bigcup$ A) /\ ($\forall$ D, D $\in$ (A $\sim$ B) -> D $\cap$ K = $\emptyset$) /\
  ($\forall$ D, D $\in$ B -> $\exists$ x, D $\cap$ K = [x]) \}.
\end{lstlisting}

According to the definition of $En\_TB$, we can define the set $T$ as follows:

\begin{lstlisting}
Definition En_T A : Class := \{ fun K => $\exists$ B, B $\subset$ A /\ K $\in$ (En_TB B A) \}.
\end{lstlisting}

On the basis of the $Property\_FinChar$, we can prove that $\bigcup n \in T$ if $n$ is a nest in $T$. Consequently, there is a maximal member of the set $T$ according to the maximal principle. Assume that $C\in T$ is a maximal member, we prove that $C$ satisfies the requirements of the theorem.
Since $C$ is a member of $T$, there is a subset $B$ of $A$ and $C\in T_{B}$. If $A = B$, then $D \cap C$ is a singleton set for every $D$ in $A$ and the theorem is proved. If $A\neq B$, then we can prove that $C\cup \{ d \}\in T$ for any $D\in A\sim B$ and $d\in D$. It contradicts with the condition $C$ is the maximal member of $T$. Overall, Zermelo's postulate is proved.
\end{proof}

\subsection{Zorn's Lemma}
Based on the maximal principle, we can prove Zorn's lemma. The description of Zorn's lemma and the statement in Coq are given below.

\begin{thm}[Zorn's Lemma]
If each chain in a partially ordered set has an upper bound, then there is a maximal element of the set.
\end{thm}

The formalization of the theorem is on the basis of Definition\,3.5, 3.6, 3.7 and 3.9.

\begin{lstlisting}
Theorem Zorn : $\forall$ (X le: Class), PartialOrderSet X le ->
  ($\forall$ A, Chain A X le -> $\exists$ y, BoundU y A X le) -> $\exists$ v, MaxElement v X le.
\end{lstlisting}

Assume that $(X,\leq)$ is a partially ordered set. In the first, we construct a class $F_{x}$ for every member $x$ of $X$. The members of $F_{x}$ are $\leq$-related to $x$. Its specific definition is shown in equation\,(4.6).
\begin{equation}
  F_{x} = \{ u : u\in X \wedge u\leq x \}
\end{equation}

$F_{x}$ will appear repeatedly in subsequent proofs and descriptions of theorems and lemmas. We formalize the set $F_{x}$ as $En\_Fs$ in Coq as follows:

\begin{lstlisting}
Definition En_Fs X le x : Class := \{ fun u => u$\in$X /\ Rrelation u le x \}.
\end{lstlisting}

Next we prove three lemmas before proving Zorn's lemma. We first give the specific content of the first lemma. The class $A$ is a chain of $X$ if and only if $\{ F_{a} : a\in A \}$ is a nest in $\{ F_{x} : x\in X \}$ and $A$ is not empty.

We first formalize the class $\{ F_{a} : a\in A \}$, $\{ F_{x} : x\in X \}$ can be obtained by modifying the input parameters of $\{ F_{a} : a\in A \}$. The specific formalization is defined as follows:

\begin{lstlisting}
Definition En_FF X A le := \{ fun u => $\exists$ a, u = (En_Fs X le a) /\ a $\in$ A \}.
\end{lstlisting}

If $F_{b}$ is a member of $\{ F_{a} : a\in A \}$, then there is a class $a$ that makes the conclusion $F_{b}=F_{a}$ and $a\in A$ set up according to the definition of $En\_FF$. In fact, $F_{b}\in\{ F_{a} : a\in A \}$ should get a conclusion that $b\in A$. So we add a property to this definition as follows:

\begin{lstlisting}
Axiom Property_FF : $\forall$ X A le a, (En_Fs X le a) $\in$ (En_FF X A le) ->
  ($\forall$ b, En_Fs X le a = En_Fs X le b -> a = b).
\end{lstlisting}

With this property, we can get a conclusion that $b$ is a member of $A$ when any member $F_{b}$ belongs to $\{ F_{a} : a\in A \}$. After defining the class $\{ F_{a} : a\in A \}$, we present the Coq formalization of the lemma as follows:

\begin{lstlisting}
Lemma LemmaZ1 : $\forall$ (A X le: Class),
  PartialOrderSet X le -> (Chain A X le <-> (En_FF X A le) $\subset$ (En_FF X X le)
  /\ Nest (En_FF X A le) /\ A $\neq$ $\emptyset$).
\end{lstlisting}

\begin{proof}
Firstly, we prove that $\{ F_{a} : a\in A \}$ is a nest in $\{ F_{x} : x\in X \}$ and $A$ is not empty if $A$ is a chain of $X$. According to Definition\,3.9, we can get $A$ is a non-empty subclass of $X$. Thus $\{ F_{a} : a\in A \}$ is contained in $\{ F_{x} : x\in X \}$ and $A$ is not empty can be proved easily. The following is to prove $\{ F_{a} : a\in A \}$ is a nest. The relation $\leq \cap\, (A\times A)$ is a total order on $A$ based on Definition\,3.9. In addition, $\leq$ is a partial order on $X$ and it satisfy the transitivity on $X$. We can prove $\{ F_{a} : a\in A \}$ is a nest on the basis of the definition of transitivity, Definition\,3.3 and Definition\,3.8.

Secondly, we prove that $A$ is a chain of $X$ if $\{ F_{a} : a\in A \}$ is a nest in $\{ F_{x} : x\in X \}$ and $A$ is not empty. According to the definition of chain, we need to prove that $A$ is a non-empty subclass of $X$ and $\leq \cap\, (A\times A)$ is a total order on $A$. The first part can be easily proved by the conditions. In the second part, we need to verify each properties according to the definition of total orders and partial orders. These properties include reflexivity, antisymmetry, transitivity and connex property. The first three properties can be proved on the basis of $(X,\leq)$ is a partially ordered set. Finally, because of $\{ F_{a} : a\in A \}$ is a nest, we can prove the connex property.
\end{proof}

Next, we prove the second lemma. If any member $y$ of $X$ is a upper bound of $A$, then $F_{y}$ contains every member of $\{F_{a} : a\in A\}$. Because of the formal definition of the upper bound, we add a condition $x\neq\emptyset$ when we formalize the lemma. The details are as follows:

\begin{lstlisting}
Lemma LemmaZ2 : $\forall$ A X le y, PartialOrderSet X le /\ X $\neq$ $\emptyset$ ->
  (BoundU y A X le -> ($\forall$ x, x $\in$ (En_FF X A le)-> x $\subset$ (En_Fs X le y))).
\end{lstlisting}

\begin{proof}
Since $y\in X$ is a upper bound of $A$, $A$ is contained in $X$ and $a\leq y$ for every member $a$ of $A$. Assume that $F_{b}$ is a member of $\{F_{a} : a\in A\}$. The relation $\leq$ is transitive in $X$ because $(X,\leq)$ is a partially ordered set. Consequently, $F_{b}$ is a subclass of $F_{y}$.
\end{proof}

Finally, we prove the third lemma. It is about the conversion between the maximal member and the maximal element. We prove that $y$ is a maximal element of the non-empty set $X$ if and only if $F_{y}$ is a maximal member of $\{ F_{x} : x\in X \}$.

\begin{lstlisting}
Lemma LemmaZ3 : $\forall$ X le y, PartialOrderSet X le /\ X $\neq$ $\emptyset$ ->
  (MaxElement y X le <-> MaxMember (En_Fs X le y) (En_FF X X le)).
\end{lstlisting}

\begin{proof}
Through the definition of $En\_Fs$, $En\_FF$, maximal element and maximal member, we can simplify the conditions and objectives of the proof. Then we can prove the above lemma by the properties of partial orders.
\end{proof}

From the above three lemmas, we can prove Zorn's lemma.

\begin{proof}
First we discuss whether $X$ is empty. If $X$ is empty, then we can get a contradiction from the definition of maximal element. Thus $X$ is a non-empty set. There is a set $\{ F_{x} : x\in X \}$ for every member $x$ of $X$. According to the maximal principle, there is a maximal member of $\{ F_{x} : x\in X \}$ and we can prove that there is a maximal element of $X$ by $LemmaZ3$. We need to prove some conditions when using the maximal principle. For each nest $B$ in $\{ F_{x} : x\in X \}$, we need to prove that there is a member of $\{ F_{x} : x\in X \}$ which contains every member of $B$. Since $B$ is a nest in $\{ F_{x} : x\in X \}$, we can get a conclusion that $A$ which is related to $B$ is a chain of $X$ based on $LemmaZ1$. What needs to be noted here is the relationship between $B$ and $A$, that is $B = \{ F_{a} : a\in A \}$. According to the conditions of Theorem\,4.6, $A$ has an upper bound $y$. Then $F_{y}$ contains every member of $B$ can be proved by $LemmaZ2$. At this point, we complete the proof of Zorn's lemma.
\end{proof}

\subsection{Well-Ordering Theorem}
On the basis of Zorn's lemma, the well-ordering theorem can be proved. We first give the description of the well-ordering theorem.

\begin{thm}[Well-Ordering Theorem]
Each set can be well-ordered.
\end{thm}

This theorem can be easily translated in Coq:

\begin{lstlisting}
Theorem WellOrdering : $\forall$ X, Ensemble X -> $\exists$ le, WellOrder le X.
\end{lstlisting}

We construct some classes and relation for proof based on known conditions. Assume that $X$ is a set. $L$ is a set consist of all the ordered pair $(Y,\leq)$, $Y$ is a non-empty subset of $X$ and $\leq$ is a well order on $Y$. The formal definition of the set $L$ is as follows:

\begin{lstlisting}
Definition En_L X := \{\ fun Y le => Y $\subset$ X /\ Y $\neq$ $\emptyset$ /\ WellOrder le Y \}\.
\end{lstlisting}

Then we define a relation $\prec$ on the set $L$ as follow: For any $(Y_{1},\leq_{1})$ and $(Y_{2},\leq_{2})$ are members of $L$, $(Y_{1},\leq_{1})\prec(Y_{2},\leq_{2})$ iff

\vskip 0.2cm
\par\setlength\parindent{0.5em} (1) $Y_{1}$ is contained in $Y_{2}$;

\vskip 0.1cm
(2) For any members $x$, $y$ of $Y_{1}$, $x\leq_{1}y$ iff $x\leq_{2}y$;

\vskip 0.1cm
(3) $Y_{1}$ is a cut of $Y_{2}$ (based on $\leq_{2}$).\\

\par\setlength\parindent{1.5em}
In the formalization of the relation $\prec$, we use the $First$ function of the ``axiomatic set theory'' formal system. We can get the first element of an ordered pair by it. The $Second$ function can get the second element of an ordered pair. The Coq statement is as follows.

\begin{lstlisting}
Definition lee (X: Class) : Class :=
  \{\ fun L1 L2 => (L1 $\in$ (En_L X) /\ L2 $\in$ (En_L X)) /\ ($\forall$ x y, x $\in$ fst(L1)
  /\ y $\in$ fst(L1) -> (Rrelation x snd(L1) y <-> Rrelation x snd(L2) y)) /\
  Initial_Segment fst(L1) fst(L2) snd(L2) \}\.
\end{lstlisting}

We want to prove that every chain of $L$ has an upper bound. Thus we assume that $K$ is a chain of $L$. Let
\begin{equation}
  Z = \{ x : \exists\, Y \leq, (Y,\leq)\in K\wedge x\in Y\}
\end{equation}

It is straightforward to formalize the equation\,(4.7) in Coq as follows:

\begin{lstlisting}
Definition En_Z K : Class := \{ fun x => $\exists$ Y le, [Y,le] $\in$ K /\ x $\in$ Y \}.
\end{lstlisting}

Next We define a relation $\leqq$ on $Z$: For any $u,v\in Z$, we randomly select a $(Y,\leq)\in K$ which make $u,v\in Y$. If $u\leq v$, then $u\leqq v$.

\begin{lstlisting}
Definition leeq (K: Class) : Class :=
  \{\ fun u v => $\exists$ Y le, [Y,le] $\in$ K /\ u$\in$Y /\ v$\in$Y /\ Rrelation u le v \}\.
\end{lstlisting}

Before proving the well-ordering theorem, we prove three lemmas first. The first lemma will be used in all subsequent proofs. We prove that the relation $\prec$ is a partial order on $L$.

\begin{lstlisting}
Lemma LemmaW1 : $\forall$ X, Ensemble X -> PartialOrder (lee X) (En_L X).
\end{lstlisting}

\begin{proof}
As can be seen from the previous definition, $L$ is a set consist of all the ordered pair $(Y,\leq)$, $Y$ is a subclass of $X$ and $\leq$ is a well order on $Y$. The relation $\prec$ is a partial order of $L$ can be easily proved by the conditions and the definition of $\prec$.
\end{proof}

The following two lemmas prove that every chain of $L$ has an upper bound. The following will be proved step by step. First we prove the relation $\leqq$ is a well order on the set $Z$.

\begin{lstlisting}
Lemma LemmaW2 : $\forall$ (X K : Class),
  Ensemble X -> Chain K (En_L X) (lee X) -> WellOrder (leeq K) (En_Z K).
\end{lstlisting}

\begin{proof}
First of all, $\prec$ is a partial order on $L$ according to $LemmaW1$. Thus $K$ is a non-empty subclass of $L$ and the relation $\prec \cap (K\times K)$ is a total order on $K$ on the basis of Definition\,3.9. We know $Z=\{x : \exists\, Y \leq, (Y,\leq)\in K\wedge x\in Y\}$ on the basis of equation\,(4.7). According to the definition of $\leqq$, total orders, partial orders and chain, we can prove $\leqq$ is a total order on the set $Z$. Next we prove $\leqq$ is a well order on $Z$. Assume that $P$ is a non-empty subclass of $Z$ and $p\in Y$ for arbitrary $p\in P$ and $(Y,\leq)\in K$. $Y\cap P$ is the subclass of $Y$, and there is a minimal element $q$ of $Y\cap P$. If $r\leqq q$ for a member $r$ of $P$. Assume that $(Y_{1},\leq_{1})$ is a member of $K$ and $r$ is a member of $Y_{1}$. If $(Y_{1},\leq_{1})\prec (Y,\leq)$, then $r$ belongs to $Y\cap P$. Thus we can prove $r = q$. If $(Y,\leq)\prec (Y_{1},\leq_{1})$, then $r\leq_{1} q$. Because of $Y$ is an initial segment of $Y_{1}$, $r$ is a member of $Y$ and $r\leq q$. Therefore, we get a conclusion that $r = q$. The above verifies that $\leqq$ is a well order on $Z$.
\end{proof}

Second we prove that every chain of $L$ has an upper bound.

\begin{lstlisting}
Lemma LemmaW3 : $\forall$ (K X: Class),
Ensemble X -> Chain K (En_L X) (lee X) -> $\exists$ y, BoundU y K (En_L X) (lee X).
\end{lstlisting}

\begin{proof}
According to $LemmaW1$ and $LemmaW2$, the relation $\prec$ is a partial order on $L$ and the relation $\leqq$ is a well order on the set $Z$. We prove that $(Z,\leqq)$ is an upper bound of the chain $K$. The proof goal can be divided into three parts according Definition\,3.6. The first part is to prove $(Z,\leqq)$ is a member of $L$. Since $Z$ is contains in $X$ and the relation $\leqq$ is a subclass of $X \times X$, the ordered pair $(Z,\leqq)$ is a member of $L$. The second part is to prove $K\subset L$. It can be easily proved by the condition $K$ is a chain of $L$. In the third part, we prove that $A\prec (Z,\leqq)$ for each element $A$ in $K$ on the basis of the definition of the relation $\prec$ and the relation $\leqq$. In summary, the ordered pair $(Z,\leqq)$ is an upper bound of the chain $K$. The $LemmaW3$ is proved.
\end{proof}

The well-ordering theorem can be proved by the above three lemmas, and the process of proof is as follows:
\begin{proof}
First, we know that every chain of $L$ has an upper bound by $LemmaW3$. Thus there is a maximal element $(Y,\leq)$ of $L$ according to Zorn's lemma. Next we will prove $Y=X$. We assume that $Y\neq X$, thus $Y$ is a proper subclass of $X$ because of $Y\subset X$. If $x$ is a member of $X\sim Y$, we define a relation $\lessdot$ on $Y\cup \{x\}$. If $y_{1}$ and $y_{2}$ are members of $Y$ and $y_{1}\leq y_{2}$, then $y_{1}\lessdot y_{2}$; if $y\in Y$, then $y\lessdot x$. Obviously, the relation $\lessdot$ is a well order on $Y\cup \{x\}$ and $Y\cup \{x\}$ is a subclass of $X$. So $(Y\cup \{x\},\lessdot)\in L$ is established and $(Y,\leq)\prec(Y\cup \{x\},\lessdot)$ by the definition of $\prec$ and $\lessdot$. This is in contradiction with the condition that $(Y,\leq)$ is a maximal element of $K$, therefore, $Y=X$. The relation $\leq$ is a well order of $X$ because it is a well order of $Y$. At this point, we complete the proof of the well-ordering principle.
\end{proof}

\subsection{The proof of AC}
In the end, we regard AC as a theorem to prove, that is to prove every set has a choice function. The formal description of the theorem is as follows:

\begin{lstlisting}
Theorem Proof_Axiom_Choice : $\forall$ X, Ensemble X -> $\exists$ $\varepsilon$, Choice_Function $\varepsilon$ X.
\end{lstlisting}

As follows, we prove the theorem according to Zermelo's postulate and the well-ordering theorem respectively.

\begin{proof}(Zermelo's Postulate)
Assume that $X$ is a non-empty set and $\tilde{X} = 2^X \sim\{\phi\}$. Let
\begin{equation}
  \mathcal{P} = \{ (A \times \{A\}) : A\in\tilde{X} \}.
\end{equation}

The set $\mathcal{P}$ is a subset of the Cartesian product $X \times \tilde{X}$. Moreover, it is a disjoint set composed of non-empty sets. The formalization of the set $\mathcal{P}$ is as follows:

\begin{lstlisting}
Definition En_p X : Class :=
  \{ fun z => $\exists$ A, A $\in$ (pow(X) $\sim$ [$\emptyset$]) /\ z = (A $\times$ [A]) \}.
\end{lstlisting}

According to Zermelo's postulate, there is a set $D$ such that $B\cap D$ is a singleton set for every $B$ in $\mathcal{P}$. Next we construct a class $\varepsilon$ as follows:

\begin{lstlisting}
Definition En_fc X D : Class :=
  \{\ fun A B => A $\in$ (pow(X) $\sim$ [$\emptyset$]) /\ B = fst( $\bigcap$((A $\times$ [A]) $\cap$ D) ) \}\
\end{lstlisting}

Obviously, the class $\varepsilon$ is a function, which satisfies the definition of relation and the uniqueness of the function. The range of it is a subclass of $X$ and the domain of it is equal to $\tilde{X}$. Finally, we only need to prove that $\varepsilon(A)\in A$ for each member $A\in \tilde{X}$. Since $B\cap D$ is a singleton set for every $B$ in $\mathcal{P}$ and $A \times \{A\} \in \mathcal{P}$, there is a singleton set $\{ (a,b) \}$ such that $(A \times \{A\}) \cap D = \{ (a,b) \}$. Apparently $(a,b)$ belongs to the singleton set $\{ (a,b) \}$ and $A \times \{A\}$.  Therefore, $\varepsilon(A) = a \in A$ is established and $\varepsilon$ is the choice function on the set $X$. The theorem $Proof\_Axiom\_Choice$ is proved.
\end{proof}

\begin{proof}(Well-Ordering Theorem)
First we consider whether the set $X$ is empty. If $X$ is empty, then we can prove that there is no choice function of $X$. If the set $X$ is not empty, $X$ has a well order $\leq$ according to the well-ordering theorem. As can be seen from the definition of well orders, any non-empty subclass of $X$ has a minimal element. Next we prove that there is a choice function on $X$.

As shown in equation\,(4.9), we construct a function $\varepsilon$ and prove that it is a choice function on $X$.
\begin{equation}
  \varepsilon = \{ (x,y) : x\in (2^{X} \sim \{\emptyset\}) \wedge y\in x \wedge (\exists z,\, MinElement \, z \, x \, \leq \, \wedge \, y = z) \}
\end{equation}

According to Definition\,3.7, we can formalize the equation as follows:

\begin{lstlisting}
Definition En_cf X le : Class := \{\ fun x y => x $\in$ (pow(X) $\sim$ [$\emptyset$]) /\
  y $\in$ x /\ ($\exists$ z0, MinElement z0 x le /\ y = z0) \}\.
\end{lstlisting}

According to the definition of the choice function, we prove that $\varepsilon$ is a function in the first step. Obviously $\varepsilon$ is a relation. If $(x,y)$ and $(x,z)$ are the members of $\varepsilon$, then $y$ and $z$ are minimal elements of $x$ according to the definition of $\varepsilon$. Consequently, the result $y = z$ is obtained. Next we prove that the range of $\varepsilon$ is a subclass of $X$. It can be easily proved according to equation\,(4.9). In the third step, we prove the domain of $\varepsilon$ is $2^{X}\sim\{\emptyset\}$. We transform the proof goal into a sufficient and necessary condition based on axiom of extent. The sufficiency can be directly proved and the necessity is to prove $A\in dom(\varepsilon)$ for every $A\in(2^{X}\sim\{\emptyset\})$. $A$ is a subclass of $X$ and $A$ is not empty. Therefore, $A$ must has a minimal element, which is denoted as $z0$. We prove $(A,z0)\in \varepsilon$ by the definition, thus $A\in dom(\varepsilon)$. Finally, according to the definition of $\varepsilon$, $\varepsilon(A)\in A$ for each member $A$ of domain $\varepsilon$. In summary, AC is completely proved.
\end{proof}

In conclusion, we have completed the formal proof of the equivalence between AC, Tukey's lemma, the Hausdorff maximal principle, the maximal principle, Zermelo's postulate, Zorn's lemma and the well-ordering theorem.

\section{Related Work}
\label{sec:5}

There already exists several formalization of axiomatic set theory. These works completed the formalization of AC, but did not prove the theorems which equivalent to AC. For instance, Werner's work \cite{W97} is to study relationships between axiomatic set theory and type theory. He has presented two families of relative consistency proofs between ZFC and the calculus of inductive constructions (CIC) in Coq. Based on Werner's work, Bruno Barras \cite{B10} has formalised the syntactic metatheory of CIC used by the Coq proof assistant, giving it a semantics in set theory and formalising a soundness proof in Coq itself. Simpson develops an axiomatization of ZFC and formalizes common set theoretical notions. See Coq user contribution \url{coq-contribs/functions-in-zfc}. Dominik Kirst and Gert Smolka formalise second-order ZF set theory in the dependent type theory of Coq in \cite{KS17,KS18}.

In addition to the above works, there are some works concerning the equivalence theorems and the applications of AC. Daniel Schepler prove Zorn's lemma and the well-ordering theorem according to AC. See Coq user contribution \url{coq-contribs/zorns-lemma}. But he uses the naive set theory in Coq standard library and only proves the two of equivalence theorems of AC. On the basis of the formal system AgdaLight, Danko Ilik prove the well-ordering theorem in type theory, instead of set theory \cite{D07}.
Moreover, Daniel de Rauglaudre present a formal proof of Banach-Tarski paradox using AC in Coq proof assistant \cite{D17}. Lawrence C. Paulson mechanized the relative consistency of AC and the generalized continuum hypothesis using Isabelle/ZF \cite{P03}.

In our previous conference paper \cite{SY17a}, we have preliminarily completed the proof of equivalence between AC, Tukey's lemma, the Husdorff maximal principle, the maximal principle and Zermelo's postulate. It is an initial version and does not give a detailed proof process of Tukey's lemma.

Our present work takes from all of the above cited works, but the axiomatic system we used is Morse-Kelley axiomatic set theory. Based on the previous conference paper, this paper presents a complete formal proof of Tukey's lemma. In addition, we improve the proof process of the Husdorff maximal principle, the maximal principle and Zermelo's postulate in this paper. Zorn's lemma and the well-ordering theorem are also proved according to the maximal principle. We finally prove AC by Zermelo's postulate and the well-ordering theorem, thus completing the cyclic proof of equivalence between AC and these theorems.

\section{Conclusion and Future Work}
\label{sec:6}
AC is an important axiom in axiomatic set theory and it has a wide range of applications in modern mathematics. In this paper, we formalize AC and several of its famous relevant theorems. These theorems include Tukey's lemma, the Hausdorff maximal principle, the maximal principle, Zermelo's postulate, Zorn's Lemma and the well-ordering theorem. We prove the equivalence between them and the proof is based on the ``axiomatic set theory'' formal system that we developed.

In this paper, all of the proofs have been implemented using the Coq proof assistant.
The complete source files containing the Coq formalization and proofs are accessible at:

\vskip 0.25cm
{\noindent \url{https://github.com/styzystyzy/Axiom_of_Choice/}}
\vskip 0.25cm

Overall, the proofs consist in about 5,800 lines of code with about 80 definitions and 100 lemmas and theorems. It has been tested and should compile under Coq 8.9.0. Table\,2 provides a detailed account of the formalization in terms of script files. To help navigation through the script files, we indicate the related sections in the paper. The count in terms of lines of code distinguishes between specifications and proofs.

\begin{table}[ht]
  \label{tab:2}
  \renewcommand\arraystretch{1.3}
  \centering
  \caption{Overview of the formal proof of AC and theorems}
  \begin{tabular}{| c | c | c | c |}
  \hline
   File                           & Reference in the paper  &  Spec  &  Proof  \\ \hline
   Logic\_Property.v              &        Section 2        &   20   &   10    \\ \hline
   Axiomatic\_Set\_Theory.v       &        Section 2        &   600  &   2000  \\ \hline
   Finite\_Character.v            &        Section 3        &   300  &   1100  \\ \hline
   Basic\_Definitions.v           &        Section 3        &   90   &   0    \\ \hline
   Tukey\_Lemma.v                 &        Section 4.1      &   90   &   340   \\ \hline
   Hausdorff\_Maximal\_Principle.v&        Section 4.2      &   30   &   120   \\ \hline
   Maximal\_Principle.v           &        Section 4.3      &   20   &   50    \\ \hline
   Zermelo\_Postulate.v           &        Section 4.4      &   30   &   270    \\ \hline
   Zorn\_Lemma.v                  &        Section 4.5      &   30   &   240   \\ \hline
   Wellordering\_Theorem.v        &        Section 4.6      &   150  &   410   \\ \hline
   Zermelo\_Proof\_AC.v           &        Section 4.7      &   10   &   120    \\ \hline
   WO\_Proof\_AC.v                &        Section 4.7      &   10   &   80    \\ \hline
  \end{tabular}
\end{table}

In the future, we will formally prove more equivalence theorems of AC. We will use Tukey's lemma to prove Tychonoff's theorem which is a famous theorem in topology. In addition, we plan to formalize ``abstract algebra'' and ``general topology'' on the basis of the ``axiomatic set theory'' formal system. It will be a meaningful exploration and attempt on the formalization of three modern mathematical structures -- ordered structure, algebraic structure and topological structure -- which are proposed by the Bourbaki group.

\section*{Acknowledgment}
\noindent We are grateful to the anonymous reviewers, whose comments greatly helped to improve the presentation of our research in this article. This research is supported by National Natural Science Foundation (NNSF) of China under Grant 61571064.

%% in general the use of bibtex is encouraged

\end{document}